# Mikhail Vasil'evich Lomonosov

## *Oration on the Birth of Metals by Earth's Tremor* [1]

*delivered at the Solemn Assembly of the Imperial Academy of Sciences on the Namesake of Elisabeth, Empress of All the Russias, on the 8th of September in the Year 1756, by Michael Lomonosow, Collegiate Councillor and Professor of Chemistry.*

[ *Oratorio De Generatione Metallorum a Terrae Motu* ]

[ *Слово о Рождении Металлов от Трясения Земли* ]

Whenever I contemplate the dreadful workings of nature, esteemed audience, I am always led to the opinion that there is not a single one of them that is solely dreadful, nor solely dangerous and harmful, which does not bring both benefit and pleasure. By some divine providence, adverse things seem to be joined with pleasant ones, so that, reasoning about adversities, we may feel greater pleasure in the enjoyment of pleasant things. We shudder at the raging waves of the boiling sea, but the winds which provoke it carry laden ships to the desired shores. Our winter's severity is intolerable to many, and heavy upon ourselves; yet it holds back the infected air, poisonous juices, and dulls the bites. Although the benefits arising from adversities often escape our notice, they are nevertheless true and significant. Thus, for countless ages, only the terror of lightning struck fear into mortals, and nothing else was feared but the scourge of an angry deity. But our age, blessed with the recent revelations of natural secrets, has given us this consolation, that through the works of natural science we got more of an understanding of the overflow of generosity rather than wrath from the heavens. Were the mountains and plains bare, devoid of the adornment of trees and plants, lacking the splendor of flowers, and the abundance of fruits, the anxious farmer would not hope for full granaries with swaying ears of golden wheat; all these benefits would have been lacking if not for the electrical energy of pregnant clouds, the fruitful rain, and a genial breeze that animates the slow growth of plants.

The truth of this matter (which had already occurred to ancient farmers, though not clearly, but had already crossed their minds) by the action of electric force, produced by the diligent hands of nature's investigators, through the acceleration of plant growth, has been so explained and demonstrated to us that there is no longer any place for doubt.

So, when such enlightenment shines forth by the revelation of natural mysteries, to our great comfort and joy, especially where previously, through the concealment of occurring pleasantness, only the image of adversity was presented before us, therefore, I consider it most fruitful to add new weight to the truth with a new argument, as far as it lies in my power.

---

[1] Translated from Latin and Russian versions (Refs. [1]) by V.Shiltsev, footnotes and commentary by V.Shiltsev.

For this purpose, I have found no more suitable subject than the movement of the earth; which, although mournful and dreadful, as we recently mourned over cities laid waste, regions desolate, and almost extinguished populations[2]; yet nevertheless, it not only contributes to our comforts, but also to our luxuries to the greatest extent, by producing, among many other benefits, countless metals useful for various purposes. I will endeavor to present this to you to the best of my ability in this present discourse, in which, after briefly outlining earth tremors, I will show the various effects they produce on the earth's surface, as well as the causes and materials serving this purpose, then the places where metals are found, and finally, how they are born.

However, this terrifying and violent phenomenon manifests itself in nature in four ways. Firstly, when the earth is shaken by frequent jolts, with a slight damage to buildings. Secondly, when it swells and rises, and then settles with an alternating and perpendicular motion; where buildings remain always safe in their position. Third, the surface of the earth, resembling waves, undergoes very disastrous oscillations, for open chasms gape upon the trembling buildings and the pale-faced people, and often swallow them wholly.

Finally, fourthly, when all the force of the tremor is directed along the horizontal plane, then the earth, seemingly taken from under the buildings, leaves them as if suspended in the air, and, destroying the bonds of their fortifications, brings them to ruin. These various earth movements do not always occur in a simple manner; but the trembling often comes with strong vibrations, meanwhile always accompanied or even preceded by underground noises, murmurs, resembling human shouts, and the sounds of clashing arms. Springs and new rivers, sometimes burst forth from the bosom of the laboring earth; and smoke, ashes, and flames often increase the terror of mortals with their immediate presence. Such frequent changes in the natural order indicate to us that the surface of the earth now has a completely different appearance from what it was in ancient times. For it often happens that lofty mountains are destroyed by the blows of earthquakes and are swallowed up by a wide gaping chasm in the earth, which is eventually filled by water bubbling up from the depths of the earth, or it is flooded by the sea rushing in. On the contrary, new mountains arise in the plains, and rising above the sea floor, they form new islands in the air. This, according to the trustworthy accounts of ancient writers and by new examples, nature has operated in all ages. Although the ancient testimonies about the changes in the earth's surface are well known to scholars, here they must find their place for the orderly arrangement of the parts of this discourse. Therefore, let us listen to Pliny (* in Naturalis Historia, *v.2*), who, from various authors, briefly recounts these changes.

He says, "Lands are born and suddenly rise from the sea: as if nature were giving some mutual payment, returning elsewhere what it has swallowed up. Long ago, the famous islands of

---

[2] References to recent major earthquakes in Peru (1745-1746) and Lisbon (1755), these events had greatly impressed he public and had been widely discussed.

Delos and Rhodes, which according to reports emerged from the sea, were born. Then the smaller islands of Melos and Anaph; between Lemnos and Hellespont, Nea; between Lebedos and Teos, Galone; between the Cyclades islands, in the forth year of the one hundred and thirty-fifth Olympiad[3], Terra and Terazia; between them, one hundred and thirty-five years later, Iera, or Automata. Then Thia, two miles away, in our time, during the consulship of Silanus and Balbus, on the first day of July; and before our time, near Italy, between the Aeolian islands; also not far from Crete, an island rose from the sea two thousand five hundred steps with warm springs. Another, in the third year of the one hundred and sixty-third Olympiad[4], in the Tuscan Gulf[5], burning with violent breath. They say that there was a great multitude of fish swimming around it, and those who used them for food soon lost their lives. The same is said about the Pithecussians, who rose in the Bay of Naples. Mount Epopeon, after issuing sudden flames, was compared to a field, on which the city sank, and another earthquake produced a lake. Mountains, previously submerged in the sea, turned into islands, which are called Prochyta. For in this way nature also forms islands. She separated Sicily from Italy, Cyprus from Syria, Euboea from Boeotia, Atlantides from Macris, and Bithynia from Besbicus, Leucosia from the Sirenian cape. On the contrary, she deprived islands of the sea and added them to the land. She connected Lesbos with Antissa, Halicarnassus with Zephyria, Miletus with Dromiscus and Pernu, Parthenian Cape of Narteucus[6]. The island of Gibanda, formerly in the Ionian Sea, now stands two hundred stadia from the sea. The Ephesus land contains a Syrian Island[7] in the middle of itself; Sophania and Derasides, nearby islands, are contained by Magnesia; Epidaurus and Orik ceased to be islands. Nature has taken entire lands, first of all, immensely extensive, where the Atlantic Ocean is, if one can believe Plato. By this, Akarnania was separated by the Ambracian Gulf, Achaia by the Corinthian, Europe and Asia by the Propontis and the Black Sea. Moreover, the sea has burst through Leucadia, Antirrhium, Hellespont, and two Bosporuses. Not to mention lakes and gulfs, the earth itself swallows itself up. It swallowed up Cibot — a towering mountain with the city of Curitum, Sipylus in Magnesia, and before that, in the same place, the renowned city of Tantalus, Galama, and Gamala — Phoenician cities with surrounding areas — and the towering Phlegyan Ridge in Ethiopia. Pyrrhus and Antissa were carried away near the Meotian Pontus[8]; Elicia and Bura as well, in whose Corinthian Gulf traces are visible in the abyss. From the island of Ceos[9], more than thirty thousand steps of land got suddenly disappeared with many people swallowed by the sea. From Sicily — half of Tindaris and all that perished from Italy, as well as from Boeotia and Eleusis."

---

[3] 232 BC
[4] 126 BC
[5] Tyrrhenian Sea
[6] Exact location is not known
[7] Territory near city of Ephesus, Western Asia Minor
[8] Azov Sea
[9] Now Kea

Such ancient accounts are confirmed by recent examples. For we see new islands born in the sea in this century. The most notable of them is in the Aegean Archipelago, near the island of Santorini. Since 1707, from March 29, during an earthquake, it began to emerge from the sea. At first, it was like a rocky hill, but over the next four years, it grew up for several miles.

However, I do not intend to present more such examples, nor to spread the poverty of the Peruvian capital city of Lima with rhetoric, nor the cruel fate of Lisbon. There is no need to further illustrate the downfall of cities by earthquakes, for the entire face of the earth is filled with clear evidence of this. Wherever you see rocky mountains with cracks, do not doubt that the remaining traces are of earthquakes, the more severe, the more chaotic the ruins, the upheavals, and the chasms.

Exploring the sufficient cause for such actions, it seems safer to me to philosophize by seeking it within the earth itself, leaving aside the opinions of the ancient Babylonians, who thought that all this originated from the power of the planets. And although Pliny presents many circumstances in their favor, and although it can be inferred from the shaking of the center (if any occurs from the mutual action of celestial spheres), to which bodies are drawn by gravity, that something can be guessed about the shaking of the earth, yet in every experiment, those things should be preferred which precede the experienced matter itself, are present with it, and follow it to completion, always closely connected with it. Therefore, for the true and general cause of an earthquake, I acknowledge - with almost all modern and ancient philosophers - the subterranean fire.

So, the spirit animating all nature presents itself primarily for consideration, which from the deepest earthy chasms throughout the surface of the earth and even in the atmosphere, displays its actions, while often being its companion. For only through many openings is it expelled, as many are the mountains exhaling fire and the chasms emitting flames. It is neither excessively strained by the heat of the torrid zone nor greatly diminished by the severity of the cold lands leaning towards the poles, but it acts everywhere and opens a path for itself in various places. Witness the fiery mountains near the equator, between the tropics, such as the Peruvian ones and those that blaze on the Indian and Green Island seas. In temperate climates, Etna, Vesuvius, Lipari, and many islands in the Aegean Archipelago, which although not with continuous heat, yet with frequent belching of flames from the very depths, clearly show that the Tyrrhenian and Aegean Seas overflow with subterranean fire. I do not mention the shores of the Caspian Sea, serving the inhabitants with hidden fire, which constantly assists them in their dwellings for cooking food and other needs after the upper earth has been removed. To the polar circles, first, the famous Mount Hekla in Iceland, then the island, Mayen, which appeared in past centuries, called "Mayen." Both places carve out a great flame, ashes, and glowing rocks amidst the eternal ice. Not far from the cold zone stand the Kamchatka ranges, breathing flames, as well as those separated from South America by the Strait of Magellan, which gave that land its fiery name. All these burning apertures clearly declare the power of subterranean fire, but they also demonstrate its effects more extensively and almost universally. For not only warm and medicinal springs, but also wells and mines dug by human labor, as well as vast seas and the great ocean itself, unquestionably indicate the internal heat of the earth. Everywhere, not only in shallow places but also in deep abysses, a great multitude of fish is found or inferred by circumstances. Wherever various kinds of whales congregate, they feed on small fish, which in turn sustain their lives with marine plants or mud. But the growth of vegetation and the softness of mud require the warmth of the sea bottom. To maintain them through so many centuries, subterranean fire is necessary everywhere, for it is highly unlikely that the rays of the sun could produce sufficient warmth at such depths through heating movement alone. Furthermore, the Arctic Ocean, covered with ice, abounds in various kinds of animals that feed on fish, clearly indicating that the sea bottom receives sufficient warmth from the internal earth fire without the rays of the sun.

Reflecting on the multitude of subterranean fire, our thoughts immediately turn to the knowledge of the material in which it is contained, requiring it to be highly conducive to ignition, impervious to extinguishment, especially in places where access to external air is difficult, and abundant throughout the entire globe. What material is more conducive to ignition than sulfur? What material is more impervious to extinguishment and better at sustaining and nourishing fire? For even when it seems extinguished, it reignites upon contact with air, as long as it is still molten and emits its vapors. What combustible material emerges more abundantly from the depths of the earth than sulfur? For not only from the jaws of fire-breathing mountains is it emitted, and in hot boiling springs and in dry subterranean chasms it gathers in great abundance, but there is not a single ore, hardly a single stone, that through mutual friction with another does not emit sulfur vapor and thereby reveal its presence within.

Some might find it astonishing that this sustenance of subterranean fire has not been exhausted over so many centuries, during which flames have been emitted through so many openings. But considering the quantity of sulfur emanating from the earth's depths, one can easily judge that the great abundance of it contained within, to which the burnt sulfur throughout the ages through combustion has a very small proportion, is as the thin shell of the earth's surface to its entire thickness.

This abundant material rightfully holds the first place among minerals, because neither plants nor animals owe any essential part of their existence to it, and it is evident that not a single metal is born without it.

You now see, listeners, the general internal sustenance of warmth, spread throughout the earth's depths, and rightfully expect me to show the very cause by which such an abundance of sulfur material ignites. To your satisfaction, I propose that through the internal movement of inert particles, composing bodies, and thus sulfur, greater friction occurs within the earth, due to its strong pressure from the bodies lying upon it, which must be all the greater the deeper the sulfur is situated; and from the strong friction of sulfur, ignition must necessarily ensue.

This fire, depending on various properties of the material lying closer to the earth's surface, may have more or less strength and erupt more abundantly for sustenance. Then, having exhausted it, it either dies or, opposed by contrary action, fades away until it gains new strength from fresh sulfur, driven by heat from the inner subterranean depths, and emits flames into the air.

Therefore, we have sufficiently understood that this warmth and fire dwell continuously in the bowels of the earth. Therefore, it is necessary to look further, if there is cold and frost, opposing them. It is true that the extensive Siberian regions, especially those lying towards the Arctic Sea, as well as those expansive fields constituting the ridge of a very high mountain, by which the Chinese state is separated from Siberia, have the ground frozen to a depth of about two or three feet throughout the summer. And although this may be attributed more to winter cold than to summer heat, overcoming it, that these places, one due to the proximity of a cold climate, the other due to the high position, have lost the gentle influence of the sky, nevertheless, not only one reason leads me to think that in some places there is a hidden cause of cold within the earth, which is capable of turning water into ice almost on the earth's surface. For, firstly, the famous Besançon cave in France (which is still considered a wonder of nature by some; it is now used as evidence of the wandering imaginary warmth-producing material or fiery element) shows us here hidden causes of action, by which a large amount of ice is produced in it, especially in summer. For, contrary to common opinion, Mr. Cossigny, through thermometer observations, assured that the dissolution of air in this cave is constant: it always shows several degrees of cold slightly below the freezing point[10]. Therefore, under the leadership of reasoning, we understand that by summer rainwater passing through the cracks of the cave's top, dripping to its bottom, and freezing into pillar-like forms. On the other hand, in winter, when water above ground turns into ice and does not enter the cave, then there is no material for the formation of ice in it. This action cannot be attributed to external air; therefore, an internal force, capable of freezing enough, must be sought. Similar to this phenomenon, I recently heard reliably that in the coasts of the Novaya Zemlya some rivers differ so much that one remains green with grass throughout the summer, while the other remains constantly covered with hardened snow, despite the fact that the sun shines equally on both riverbanks, from which it can reasonably be inferred that the interior of the banks, due to the difference in subterranean warmth and cold, shows this difference.

Such phenomena are consistent, it seems, with the following reasoning, which I hope will be sufficient to understand the cause of subterranean cold. We have seen above that not only cities and islands, but entire lands are sometimes swallowed up by earthquakes. Therefore, it is not surprising that if places lying near the poles or the tops of ice-covered areas were closed by earthquakes into the depths of the earth in ancient times and, being buried by a great multitude with ice and snow, they no longer feel the warmth of the sun. Art and common people have taught to preserve ice in cellars throughout the summer, which rarely occupies more than twenty cubic fathoms. How long will it take for such a quantity of ice, which contains several million cubic fathoms within itself, to melt in the earth's depths?[11] Truly, many centuries must pass before the excess of its frost communicates with the earth's bowels touching it, comes into equilibrium with it, and finally melts, turning into water from subterranean heat. How long it takes to complete this task of nature! Not only individual species, but entire nations may arise and perish. This, though probable, no one will deny that subterranean fire is much stronger than that frost, because it comes from the earth's surface and is the fruit of the cold external air; fire, on the contrary, reigns in its homeland.

---

[10] Jean-Francois Charpentier de Cossigny, "*Extrait d'une letter écrite de Besançon a M. de Réaumur le 29 novembre 1743, sur la grotte qui se trouve a quelque distance de Besançon at qu'on en nomme la glaciére*", in "Mémoires de mathématique at de physique, présentés á l'Académie royale des scinces par divers savant et lus dans les assamblees", v.I, 1750, pp.195-211.

Following this abundant and inflammable mineral sulfur are those materials which have their origin from growing and animal bodies and, upon entering the earth's depths, have participated in minerals. Of these, mountain salt deserves the first place, which, although usually classified among minerals, owes its birth to growing and animal bodies. This, to prove briefly here, I must first establish that all mountain salt is sea salt; secondly, that sea salt originates from the decay of plants and animals.

Beginning this, I recall that in mountain salt marine animals are found, (* Ulysses Aldrovandi, in the *Museaum Metallicum, Book 3, Chapter 3*)[12] clearly indicating that it was previously liquid, meaning it was dissolved in a large quantity of fresh water, making it passable for animals. Moreover, mountain salt mostly consists of grains of various sizes, cubic in shape, as commonly occurs when sea salt settles during evaporation, which undoubtedly proves that mountain salt, from brine, after the evaporation of excess water moisture, settled into a granular form. The larger and harder parts usually settle more from the brine, and the longer the evaporation, the more pronounced this effect is. Such natural chemical action from the shaking of the earth can easily be understood. Let us suppose that from the bottom of the sea (as sometimes happens) an island with a sandy valley in the middle emerges and lifts it above sea level, filled with brine. In such circumstances, who would doubt that fresh water, partly filtered through sand, partly evaporated into the air, must leave salt in its dry form; which later, with sand flowing from the mountains or earth, or with sand and ash from fiery mountains, can be covered. Therefore, when the salinity of the sea does not come from mountain salt, as many thought, but conversely, this is more likely to be the case, another origin of sea salinity must be sought.

---

[11] This description of the permafrost phenomena and explanation of extensive periods of time needed for its thawing is seemingly the first in the scientific publications of 18th century; somewhat earlier, in 1684, the military governor of Yakutsk, a town along the Lena River, reported to Russian tsars Peter and Ivan that he was able to dig a deep well because "the earth in summer thaws one and a half arshin [0.71 m], and it never thaws by more than two arshin [1.42 m]"; see also Lomonosov's "*On the Earth's strata*" ([1], v.5, p.567).

[12] Ulysses Aldrovandi, *Museum Metallicum* (Bologna, 1748).

The effort that many have futilely made on this issue is facilitated by the chemical separation of the salt mixture. For it is known that sea and mountain salt consist of alkali and acidic spirits[13]. The alkali salt, which constitutes sea and mountain salt, is the same, derived from the ashes of various trees, that is, potash, and differs only slightly by the addition of chalky or lime material. The acidic spirit is mixed with general acid with an added mercury or arsenical original material. Of all the salt there is in the world, I affirm that, mixing from alkaline and acidic materials, resulting from the decay of plants and animals, it has multiplied over time to such abundance. But here arises the question for me, where does such a multitude of alkali and acidic materials come from, so that there are enough of them to make up all the salt? However, I also rightly ask the opposite: where would such a multitude of alkali and acidic materials go, which are born in countless quantities every day, if the vast seas did not take them into their extensive depths? For if we could directly count everything, as much alkali and acidic material is born from burning many trees and herbs, as many by fires, various buildings in cities and villages, fires in vast steppes and forests every year, or better to say daily, growing things turn into ash on the entire surface of the earth, and how much alkali from the ash is washed away by rains and rivers into the sea—then we would acknowledge that all seas must already be alkali. But by the wise, divine gaze, this corrosive material is softened and, by uniting with another, becomes suitable for common use. For, although through the burning of growing things much alkaline material is born from them, nevertheless, a sufficient number for saturation in the mixture of the first and for the formation of salt is given to us by the decay and putrefaction of animals and growing things, from which the former produces volatile acid, and the latter produces arsenical material required for it, which, if it should be abundant, can be judged from the multitude of trees, leaves, and herbs, as well as animals, being destroyed by putrefaction and decay throughout the entire surface of the earth, from which the mercury original material is separated from the mixture. I omit here the Second place is occupied by underground rich materials, such as slate, coal, asphalt, mineral oil, and amber. From the following, it is evident that they owe their origin to plants. For slate stone is nothing else than black earth, born from the decay of grass and leaves, which, in ancient times, washed from fruitful places and forests by rain, settled like mud at the bottom of lakes. Then, as they dried up or were covered with sand, they became stone through long-lasting old age. It is not surprising that traces of grass and bones of river and lake fish are found in slate. Coal is associated with burnt trees, which sometimes turn out to be incinerated, and also by burning itself with ashes and potash given off, and through distillation produces bitter oil, similar to resin. The mountain oils and resins proclaim themselves with lightness and bitterness that they are of the same origin. Their birth from fossil coal can be produced, which, from its extensive layers, emit various liquids and colors through the power of underground fire, for the reception of various adjacent minerals, such as asphalt, oil, mineral oil, which differs so little from turpentine oil (distilled oil from resinous trees) that one is inadvertently taken for the other or is sold with an admixture. As for amber, it is not possible to admire enough that some learned people, with great names and merits, recognized it as a genuine mineral, regardless of the countless small creatures enclosed in it, which live in the forests, nor on the many leaves visible inside the amber, which all seem as if alive, and truly testify against that opinion and genuinely proclaim that these creatures and leaves once adhered to liquid resin, exuding from trees, then the same from above poured over them and enclosed remained. By what means they came into the earth, he will hardly understand, who has no knowledge of such great changes in the earth's surface, as we have seen above. Moreover, amber is found in Prussia under a layer of rotten wood, which, as it seems, rotted due to its age; meanwhile, the resinous material, resisting decay due to its oily nature, survived with the enclosed creatures and finally became harder underground over time due to mineral juices.

---

[13] Acidic spirit – hydrochloric acid, HCl; at then time's chemical theory the acid comprised "a general acidic matter" – what we now call hydrogen ion – and either mercury or arsenicum – "pungent original matter" (chlorine was not discovered yet).

But this is enough about fatty mountain materials. Let us finally imagine the fossilized animal bodies, which astonish many, so that they cannot convince themselves that they were ever truly animals, but were fabricated by the play of extravagant nature.

But those who do not consider nature so playful, that they might exclaim, like Narcissus:

> *…Crudelis tu quoque falsis*
>
> *Ludis imaginibus…*
>
> *["Cruel one, why do you appear before my gaze*
> *And enchant me with false visions?"]*

but by the most certain criterion of the animal kingdom, the empyreumatic oil, extracted by distillation from petrified substances, they are convicted, they confess that these true animals, mainly raised from the sea bottom by the movement of the earth, buried, penetrated by stony sap, have hardened.

These are the most notable bodies, which are content with interpreting the birth of metals. It was deemed beneficial to prove their origin, so that it would be clear how many parts of plants and animals serve in the birth of metals. Therefore, now it's time to show the places where metals are found. These are considered four main ones. First, ore veins, which are nothing else but fissures in the mountains containing various minerals and ores. Their position varies almost infinitely according to the different sides they extend to and the angle they incline towards the horizon. Second, layers in the mountains are horizontal. Third, nest-like ores. Fourth, those found on the earth's surface, such as sand containing gold, tin ores in England, bog and field iron ores, which are abundant in Russia, Sweden, and Finland. All these metal treasures are prepared by the shaking of the earth, which must be presented here. But before all else, it is necessary to see what horizontal layers and veins are like and how they are produced.

When wells are dug, various layers are revealed. Examples of this often occur, but it is regrettable that they are very rarely described. Therefore, raise, listeners, your mental gaze to the banks of the great rivers, by which the Russian state is especially nourished, where among many things worthy of attention, these steep banks are presented, which have their origin from the erosion of the undermining water. What a wonderful sight the various layers attract human sight! There, various colors are visible, sometimes different hardness and composition of the earth's interior; there, layers of fallen forests and earth deeply covered, sometimes the bones of animals and wooden works of human hands penetrate from the midst of the crumbling earth. All these spectacles are of such a nature that hardly anywhere does nature reveal her underground secrets more than in these steep banks. Of these layers, those that belong most to my business consist of sandstone or limestone, as well as shale, coal, and fossilized wood, hiding ores of various metals within them. Such layers are found in abundance in mountains rich in metals. In Germany, the most famous among them is in the Landgraviate of Hesse near Frankenberg, which contains copper and silver. There I happened not without wonder to see not only trees but also whole bundles of petrified wood containing copper and silver ore, so that in some ears of grain grains were covered with pure silver like thin beaten wire. Such horizontal layers in the rocky mountains intersect and end with metallic veins, which, although they extend into the earth in various lines from above, all widen at the bottom, compress upwards, so that they are almost completely closed on the surface and lie under black soil or other superficial soil. This kind of vein is the main and constant one. Furthermore, it is noted that such metallic veins are more often found in gentle mountains; very high and steep mountains rarely contain such wealth. And although they sometimes appear, they are always unstable, they do not run continuously through a whole mountain, but, intersecting, deprive miners of hope of acquisition. As for the material with which the veins are filled, the first place is occupied by stones, different from the rest of the mountain, such as quartz, quartzite, spar, and others. All these veins are produced by the shaking of the earth, which is confirmed by the following proofs. Firstly, the greatness and strength of the shaking vary according to the size and shape of the mountains. For the stronger the cause and the less resistance from the ground above, the greater the shakes and the strongest follow the actions. Ignited, a large amount of sulfur in the earth's depths expands and heavy air in the abysses pushes it into the overlying earth, raising it and producing different movements in different directions with different amounts of motion; and in those places, it breaks through first where it finds less resistance; the broken earth's surface shoots light parts into the air, which, falling, occupy the surrounding fields; the rest, overcoming the heaviness with its great size, forms a mountain. For the fields shaken by such force do not return to their former position, but, like disordered ruins, collapse, leaving hollow places in the gaps. From this, huge heaps rose higher than the rest of the earth's surface, belching smoke, ash, and sometimes flames with hot stones. Others, after the fire is extinguished, from ancient times with hollow interiors, resound. But while their depths still burn incessantly or intermittently with fire, at that time they throw out a great amount of various materials onto the surface, of which we have testimonies left by many writers, by which sandy and stony floods left us a memory. Cicero writes (*De Natura Deorum, v.2)[14]: "Let us think of such darkness as it was, according to the report, which the eruption of Etna darkened the surrounding lands, that through two days a man could not see a man." Such dark and dense clouds of sand and ash, falling to the ground, covered so many plants, suffocating them! Borel writes about the eruption of Etna in 1669: "Then, for three whole months, ash constantly fell, like rain, in such quantity that it covered all the surrounding fields for fifteen miles and lay so thickly that grapevines and bushes were closed off by it."[15] It takes a long time to calculate such fiery floods, by which not only Etna and Vesuvius often inundated nearby places, but also new mountains, such as the one that rose near Pozzuoli in 1538, emitting sand and ash with flames. Based on all these actions, we are quite confident that many bodies, adorning the earth's surface with such dry underground rains, are buried. Whole forests are covered, ignited by hot stones. Cornelius Severus writes (*in the poem called *Etna*):

> *Nam quando ruptis excanduit Aetna cavernis,*
> *Ardebant arvis segetes et mollia culta*
> *Jugera cum dominis, sylvae, collesque virentes.*
> *[For when Etna, with its ruptured caves, blazed forth,*
> *The fields were burning, the soft crops,*
> *Together with their masters, the woods, the green hills.]*

From such actions it's not surprising that within the earth we find layers, in which plants, not only combined with minerals, but also turned into stone, are visible. For beneath the mountain, as previously shown, and after a long time from sand, ash, and sulfuric matter solidified, they themselves can petrify and produce those ores. And extinguished trees and other plants are sometimes detached either as hardened coal or as ores. For when rainwater permeates the mountains, it carries the finest earthy particles, from which stones settle, and gains the strength to transform other bodies into stone, leaving in their crevices those particles which it had previously taken from the rocky mountain. This is evidenced by many caves and mineshafts, in which dripping water leaves accumulated stone along the walls and vaults.

It's already clear to you, listeners, the appearance, substance, and genesis of horizontal layers containing ores and other minerals within themselves; likewise, you have understood sufficiently that for the production of these, strong earth tremors and belches from fiery-breathed mountains of various underground bodies are required; therefore, let us now proceed to the origin of veins containing metals.

When the fields and forests, already extinguished by sand, ash, and stones from fiery-breathed mountains, shall cease, then with the passage of time, the smoldering, remaining fire of hidden matter sometimes endeavors to reignite the flames; from the elasticity of expanded air, the earth, rising and falling, moderately trembles, emitting through crevices the foul stench of heavy smoke, which sometimes ignites into flames. The decaying combustible matter in the piled horizontal layer compresses, while the weight above presses down, crushing the layer. Hence, gentle mountains and valleys are born, with fissures extending in various directions, intersected, from which the main ones reach from the top to the horizontal layer, while the others are cut off or disappear. When this happens, the lower convex side of the descending sedimentary land must widen the fissure, leaving the upper parts narrow. Hence, it is evident why veins are wider towards the earth's center and narrower upwards, so that they rarely appear on the surface. Meanwhile, rainwater seeps through the interior of the mountain and carries dissolved minerals with it, entering into those fissures either by squeezing or dripping; it leaves behind such a quantity of stone material in them that it fills all those cavities in some time. This is confirmed by the everyday art of miners, who very often find new minerals in exhausted mines, with which not only the broken old ores collected in heaps grow together again, but also old mines are filled with new material.

---

[14] Cicero, *De Natura Deorum, v.2, ch.38, paragraph 96.*
[15] G.A.Borelli, *Historia et meteorologia incendi Aetnae anni 1669* (Redgio, 1670)

In addition to the mentioned settlements, caused by moderate tremors, which open up fissures in the mountains for mineral veins, there are also the phenomena of elevation and subsidence, imperceptible over time. This is not only observed on the earth's surface but also clearly demonstrated in the earth's depths in mines. For empty crevices, through which intersected veins are shifted sideways, as well as gaps through which veins are separated from the mountain and composed of different materials from both, vividly illustrate that they were born after the formation of the veins by their significant expansion, when the earth had sunk even lower.

These places of both kinds, containing metals within them, originate, as is already evident, from seismic activity; a third kind must undoubtedly be attributed to the same cause. For when carefully examining the clusters of ores found nested amid the mountains, it can be inferred from the connection of stones to them from the very mountain through the aforementioned mineral gaps, that they are nothing but disrupted veins caused by a new strong tremor, resulting in their disorderly arrangement.

The fourth kind, constituting ore-bearing sites where metals are found on the earth's surface, may originate from tremors, although there may be doubts about it, yet arguments can be presented which must resolve them. For all the gold found in small grains on the surface is washed out from pure or mixed with earth sand. Physicists unanimously agree that sand originates from crushed stones. Therefore, no one would consider it impossible that the golden grains were forcibly detached from a vein and scattered among the sand by the violence of nature. To this is added the force and significance of quartz stone fragments, fused with the golden grains found in the sand, clearly indicating that sand gold is born in veins. For veins containing pure gold are almost always composed of quartz. As for the ores of tin in the Aglin mines, one should reason no differently than about bog iron ores, that they are washed out of veins penetrating the mountain by rainwater and flow into marshy valleys. But as the mountains and veins, as we heard earlier, have their origin from seismic activity, therefore the mentioned gold, iron, and lead ores must also have their birth from the same cause; consequently, all places where we see metals are produced by the shaking of the earth.

Having explained all this in order, it is necessary to show how metals are born in layers and veins, and how seismic activity contributes to their precise formation. In embarking on this, I encounter a recurring question: are metals continuously born or were they created along with other things since the world's creation and exist in the same quantity, only being squeezed into layers and veins from the bowels of the earth, where they are scattered? We have many arguments from both sides; however, the dispute will not be entirely resolved until a considerable quantity of some metal is produced by chemical experiment from non-metallic substances or until one metal is transformed into another without any trickery or error, and clearly demonstrated by chemical analysis. Indeed, there are testimonies from credible individuals who claim that silver can be transformed into gold through numerous smeltings and quenchings. Such experiments and others like them would have forcibly compelled us to agree with this opinion if they could have been demonstrated conveniently. For the artificial birth or transformation of metals achieved through artistry would serve as proof of their natural occurrence. Therefore, setting aside such speculations that often lead into dark alchemical labyrinths, and being satisfied with just one argument of similarity, I concede to the side that asserts that metals are still born today. For, based on evidence from numerous chemical experiments, metals are indeed mixed bodies[16], which implies that the substances mixed into them must undoubtedly have had their existence in nature before being blended into metals. It is difficult to imagine that all these mixed substances would have completely amalgamated during the initial production of metals, leaving nothing for subsequent times. But let us examine the birth of metals themselves in mines and veins: by some signs, we may determine which opinion should be favored.

Firstly, according to the consensus among miners, it is known that in mines there are certain veins, antagonistic to sulfuric and arsenical vapors, which impregnate the stone material growing on the walls and extracted from the mountain with water, causing it to harden and acquire a metallic luster, thus earning the name of ores. When subjected to the action of fire in the smelter, it emits vapors that settle as sulfur and arsenic in the flues and chimneys. The solid remaining part, when subjected to intense heat, yields various metals. It often happens that ores, even while still in the ground, emitting vapors or flames resembling lightning, turn into dust, from which no further metal can be obtained through smelting. Such places are called "dead metal" by miners, and when they find such in the veins through their labor, they often say the proverbial phrase: "We've come too late!"

---

[16] Lomonosov conveys a common 18th century view that metals consist of earth or glass (oxides) and combustive matter, see also "De tincturis mettalorum" ("On the luster of metals") ([1], v.1, pp.390-417).

When discussing such phenomena, reason turns between two opinions, unsure whether metals exist in a state of mixture or if they wander through the hollow subterranean abysses with the mixed-in substances separately. It wouldn't be contrary to reasoning to assert the former if these changes occurred at such depths where the air, compressed by the weight above, was halved or even tripled in pressure, causing bodies, which are constant in fire, to become volatile, or if the heat there were as intense as required to drive arsenic and sulfur with their accompanying metals into the air. However, since the aforementioned phenomena occur in places not only deep but lacking in great heat, it must be assumed that it is not the metals themselves in their entirety but the necessary substances for their mixture that fly separately to be mixed. For it is known that when arsenic and sulfur are driven upwards by fire, especially when they must carry the weight of metal with them, they become volatile. Therefore, those vapors which traverse the cavities of the mountains must be much finer than arsenic and sulfur. The substances constituting them, which are the same as those of which metals are composed, are more predisposed to mixing, as evidenced by their easy blending in fusion and other chemical experiments. When volatile sulfuric acid and its combustible matter are present, sulfur manifests when it is destroyed by flame. Arsenic consists of fine earth mixed with acidic spirit, resulting in its volatility, as evidenced by its resemblance to sublimate[17]. The said acidic spirit, combined with combustible matter, is volatile and flammable, producing phosphorus from them.

However, this has already been explained more simply and conveyed to the learned world by me before (* *Novi Commentarii, Tom. II* [18]). Let us now proceed to the general types of ores extracted from mines. Firstly, metals come out of mines combined with other minerals and are called ores, or they emerge as pure ores without any foreign matter. Ores manifest in two distinct forms: some maintain their inherent regular shapes, such as cubic marcasites, yellow spherical galena, angular white galena, needle-like antimony, and many others. Pure native metals rarely exhibit crystalline forms, although I have observed gold and copper in angular fused chunks. Mountain crystals—greenish and soft—were attached to the copper. Other ores and the majority of them lack any consistent shape and emerge as mere mixed material, such as white and red silver ores, yellow sulfuric galena, and almost all iron-bearing stones.

These four types have the following causes: metals, mixed in ores due to an disproportionate amount of mixed-in materials, expelled the excess from their mixture, from which sulfur, arsenic, and other minerals were born upon separation. Pure native metals were separated through the action of the chemistry of nature. It is evident from this that in ore-bearing places, only those metals are found pure, which descend in their pure form from solutions through chemical artistry, namely gold, silver, copper, and mercury. Besides these, neither pure metals nor semimetals are found in the earth, nor can they be returned to their original form through artistry. Copper and silver are separated from arsenic by the required heat; the latter sometimes remains in heaps like fine wires, adorned with various colors, which are the traces of expelled arsenic. This is achieved through prolonged exposure to fire, which, to drive arsenic into the air without excess, is necessary, pulling it into threads behind it. A marvelous concordance of art with nature! Other metals are never observed to be drawn into such thin strands, except silver and copper. The crystalline shapes in which ores and sometimes pure metals are found have a similar origin to various kinds of salts. Firstly, dissolved in water, they seep into the cavities of the mountains, where prolonged desiccation causes them to settle, a process similar to their disposition in crucibles with salts. Ores and metals without definite external shapes are born simply through mixing, like ordinary chemical substances.

---

[17] Sublimate – mercuric chloride
[18] The paper cited by Lomonosov is published not in *Novi Commentarii*, but in *Commentarii Academiae Scientiarum Imperialis Petropolitanae* (t. XIXm pp.286-298), and is titled "De tincturis mettalorum" ("On the luster of metals") ([1], v.1, pp.390-417).

Finally, it remains to be shown from where these substances, which come into the earthly clefts and produce the aforementioned actions through their mixing into ores and metals, originate. There is no doubt that the finest combustible and acidic matter is separated from sulfur destroyed by underground fire. Some attention is required for arsenic, which, combining with earth, forms semimetals, to which metals are variably related. However, the truth will soon be evident as we consider the immense quantity of hidden underground salt. For by the action of internal fire, alkali matter combines with earth or stone, releasing acidic spirit, which, upon separation, emerges into the clefts.

Thus, it has already been demonstrated that many living and thriving entities contribute to the birth of metals. This is further corroborated by the fact that petrified marine animals mostly exhibit arsenical pyrites within them, presumably from the saline matter in the sea. Also, ore veins are richer in the intermediate depths, becoming poorer the deeper they go, as if by approaching the earth's surface, they receive more vapors from animals and plants, thus producing more abundantly. However, this is freed from all doubts by the return of metals to their former inherent state from destruction when, by adding coal to their ash or glass and alloying, they regain their metallic brightness and flexibility. Metals that have arsenical matter in their mixture require coal for their return to a metallic state, which is akin to the same matter, namely from the burnt fatty parts of animals.

A vast field still remains where the mineral kingdom presents countless bodies and phenomena in the depths of the earth for contemplation, the detailed examination of which is not pertinent to my endeavor. However, a brief summary of all this discourse will suffice for its conclusion.

We have seen, listeners, a great multitude of fires in the bowels of the earth, and the abundance of sulfur necessary for its sustenance, sufficient for earthly tremors and the production of great, calamitous but also beneficial, terrifying yet also pleasurable changes. We have understood that the consumed parts of animals and thriving entities contribute to the birth of metals, whose beauty serves magnificence, hardness serves durability, and rigidity serves protection. But in your thoughts, a dreadful sight of the trembling face of the earth turns! Turn away, turn away your mental eyes from that, and carefully consider the upheavals raised by the trembling earth, with its cooling and healing sources flowing from them, gathering into rivers for our refreshment and serving the needs of various human desires.

Look upon your blessed homeland and compare it with other countries. You will see in it the moderate action of subterranean fire by nature. Not raised by Alpine or Pyrenean severe peaks to eternal winter dominating the upper atmosphere, nor lowered by deep chasms into marshy dampness, our lands are adorned with gentle ascents and slopes of fruitful fields, not devoid of metals, spreading towards our delight. Not disturbed by frequent earthquakes, scarcely heard of here, but enjoying the internal tranquility of the earth and the whole society. Oh, how blessed is Russia with these qualities! But it is made a hundredfold more blessed by the unparalleled virtues of the great Elizabeth! For under her most auspicious reign, not only do illustrious achievements in military and civil affairs and new inventions prosper with divine favor, augmenting the happiness of the homeland and captivating the admiration of the whole world, but even nature herself seems to respond to her noble qualities by showering us with her gifts. Besides the treasures enclosed in the bosom of the earth, Russia boasts of the abundance of crops and ascribes all this to the happy rule of her autocrat, having rendered thanks to the Supreme Creator. Especially on this holiday, it celebrates its state in harmony with her name; and looking across Europe at the flames of war[19], it speaks with the united voice of its subjects: "Beyond my wishes, you are solicitous for my welfare, great Empress! Rich, adorned, illustrious, and securely protected from all sides, I rejoice. I hear the resounding of your victorious weapons with full confidence. Now that arrogant enemy, having attacked your faithful allies unexpectedly, is compelled to retreat, having experienced your invincible strength. By the favor of heaven, your authority, power, legitimate counsel, and favoring fortune, your intention will be fulfilled with success. And by the glorious victories over your enemies, the Creator of the Universe, who spread the waters over the surface of the Earth and thus tamed the fearsome fire within it, will quench the flame of war with the rain of grace and pacify the world with your peace-seeking army."

---

[19] Reference to then just started the Seven Years' War (1756-1763), a global conflict involving most of the European great powers and fought primarily in Europe and the Americas.

# ORATIO
## DE
# GENERATIONE METALLORVM
# A TERRAE MOTV,

HABITA

IN SOLEMNI CONVENTV

QVO

ACADEMIA SCIENTIARVM IMPERIALIS

DIEM LVSTRICVM

# ELISABETAE AVGVSTAE

AUTOCRATORIS OMNIUM ROSSIARUM

CELEBRAVIT

IIX. ID. SEPT. ANNO MDCCLVII.

AVCTORE

*MICHAELE LOMONOSOW*

CONSILIARIO et PROFESSORE CHYMIAE.

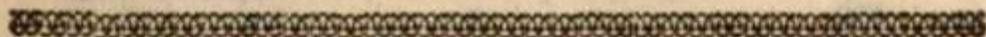

*PETROPOLI*
TYPIS ACADEMIAE SCIENTIARVM.

*Commentary* [by V.Shiltsev]:

This English translation of Mikhail Lomonosov's seminal work is derived from its Russian and Latin originals [1] (also Ref.[2], pp.295-347). It continues the series of English translations of Lomonosov's nine most important scientific works, which were included by Lomonosov himself in the convolute *Lomonosow Opera Academica* sent for distribution among Academies in Europe. Four other translations of these publications can be found in Refs. [3-6], while translations of "Oration on the Use of Chemistry", "Oration on the Origin of Light" and "Meditations on the Solidity and Fluidity of Bodies" are included in Henry Leicester's book [7]. More on the life and works of the outstanding Russian polymath and one of the giants of the European Enlightenment can be found in books [8, 9] and recent articles [10-16].

This "*Oration…*" was first published as a separate edition in 1757: the Russian publication was shortly followed by the Latin translation entitled "Oratio de generatione metallorum a terrae motu, habita in solemni conventu quo Academia Scientiarum Imperialis diem lustricum Elisabetae augustae autocratoris omnium Rossiarum celebravit IIX. Id. Sept. Anno MDCCLVII. Auctore Michaele Lomonosow, consilario et professore chymiae. Petropoli, typis Academiae Scientiarum [1757]". The "*Oration…*" was read at the celebratory public meeting of the St.Petersburg Academy dedicated to the namesake day of Russian Empress Elizaveta Petrovna (1709-1761).

Such public presentations were of great importance the Academy reputation at the Imperial Court and, therefore, its financial and political support. Correspondingly, the speakers and the topics were extensively discussed by the Academy. At the Conference of the Academy of Sciences on May 21, 1757, Lomonosov proposed several topics for upcoming 1757 celebratory meeting. "The colleagues deemed more worthy…" three topics, including the theme "De motu terrae metallifero" (On the Metal-Bearing Movement of the Earth). Lomonosov promised to develop one of these themes and "present to the Conference a dissertation written in Latin and Russian" (*Conference Protocols*, II, pp. 380–382). At the meetings of the Conference on August 19 and 20, Lomonosov read the prepared speech "Oration on the Birth of Metals by Earth Tremors" (*Conference Protocols*, II, p. 388). On August 22, 1757, "the speech by Counselor Lomonosov, read at the previous Conferences, was approved by the majority of attending academics, who attested to this with their own signatures" (*Conference Protocols*, II, p. 388). On September 6, 1757, Lomonosov read the "*Oration…*" at the solemn meeting of the Academy. At that time, this speech was also published in Russian, and shortly thereafter in Latin.

This translation was initially composed with the aid of Google Translate and ChatGPT programs, which provided a rough draft from both Russian and Latin versions of the Lomonosov's "*Discourse…*". However, the machine translations were found to be unsatisfactory as they often failed to capture the nuances and complexities of the original text - mostly due to very difficult old Russian language of the Lomonosov's original publication, which is not always easily readable and understood even by the modern day Russians. Therefore, extensive reworking and improvements by the translator were indispensable to ensure the translation accurately conveys the intended meaning and is free from semantic deficiencies.

Lomonosov (1711-1765) first delved into the mining and geological processes during his studies in Germany in 1736-1741, where he was sent as one of the most gifted students at the St. Petersburg Academy's University [16]. Upon his return to Russia, he immersed himself in research in this field until his passing. Notably, volume 5 of Lomonosov's Complete Works is entirely dedicated to his contributions to mineralogy, geology and geophysics, metallurgy, and the mining industry, encompassing 20 publications. Mikhail Lomonosov was the first Russian scientist deeply intrigued by the physical processes occurring within the Earth's major layers. On this subject, he articulated many fascinating thoughts and hypotheses, created a series of original geophysical instruments, including the first gravimeter, and conducted field research, particularly in the atmosphere. He formulated a range of geophysical problems concerning weather and climate, ocean currents, earthquakes, tectonic activity, spatial and temporal variations in the Earth's magnetic field and gravitational field, some of which were only resolved in the late 19th century. Besides this "*Oration…*", two other highly referenced works by Lomonosov in the fields of mining, geology, and geophysics are "First Principles of Metallurgy or Ore Mining" (St. Petersburg, 1763, [2], v.5, pp. 397-631), and its appendix "On Strata of the Earth" [18].

In these works Lomonosov highlights the role of water and wind in shaping the Earth's features, distinguishes between external and internal forces shaping the surface and concludes on the predominance of internal forces in the formation of mountains, volcanoes, tectonic activity, and the origin of earthquakes, carefully substantiating his conclusions with numerous examples.

Of a special note are Lomonosov's views on the "Earth tremors", the cause of which he saw in the redistribution of the Earth's deep heat. Subdividing earthquakes into fast and "insensitive long-term" ones, to which he attributed slow century-long oscillations of the Earth's surface, manifested in the birth of mountains and valleys, in the inclinations of Earth's layers, in ruptures and shifts of ore veins, he distinguishes four possible types of them (see above): "… Firstly, when the earth is shaken by frequent jolts, with a slight damage to buildings. Secondly, when it swells and rises, and then settles with an alternating and perpendicular motion; where buildings remain always safe in their position. Third, the surface of the earth, resembling waves, undergoes very disastrous oscillations, for open chasms gape upon the trembling buildings and the pale-faced people, and often swallow them wholly…Finally, fourthly, when all the force of the tremor is directed along the horizontal plane, then the earth, seemingly taken from under the buildings, leaves them as if suspended in the air, and, destroying the bonds of their fortifications, brings them to ruin. These various earth movements do not always occur in a simple manner; but the trembling often comes with strong vibrations, meanwhile always accompanied or even preceded by underground noises, murmurs, resembling human shouts, and the sounds of clashing arms."
In this passage, modern day seismologists can easily discern the astute description of earthquake mechanisms, their oscillatory nature, and the differentiation between longitudinal and transverse waves. It's worth noting that Lomonosov's "*Oration...*" predates Michell's paper, traditionally credited with the discovery of the wave-like movement of the Earth's surface during earthquakes, by three years (1760)[20].

However, it's important to acknowledge that some of Lomonosov's views expressed in this "Oration..." are either wrong or may look unconventional by today's standards. For instance, his assertion that sea salt originates from the decay of plants and animals seems antiquate(see above). These instances underscore the non-linear path to scientific truths, often requiring the exploration of numerous hypotheses before arriving at an accurate description of natural phenomena.

---

[20] In 1760 English natural philosopher and geologist John Michell published in the *Philosophical Transactions of the Royal Society* LI (1760) "Conjectures concerning the Cause and Observations upon the Phaenomena of Earthquakes" where suggested that earthquakes were experienced as seismic waves of elastic compression travelled through the Earth.

I would like to thank Prof. Robert Crease of SUNY, my long-term collaborator and co-author of several scholar papers on Mikhail Lomonosov, for the encouragement to translate Lomonosov's major works to English.